\documentclass[preprint2]{aastex}

\usepackage{epstopdf}

\newcommand{\ob}{$\Omega_{\mathrm{b}}$}
\newcommand{\obh}{$\Omega_{\mathrm{b}}{\cdot}h^2$}
\newcommand{\deu}{D}
\newcommand{\tro}{$^3$He}
\newcommand{\qua}{$^4$He}

\newcommand{\sep}{$^{7}$Li}

\newcommand{\hli}{$^4$He, D, $^3$He and $^{7}$Li}

\newcommand{\lap}{\mathrel{ \rlap{\raise.5ex\hbox{$<$}}
	            {\lower.5ex\hbox{$\sim$}}  } }

\shorttitle{BBN after Planck}
\shortauthors{Coc \etal.}

\begin{document}

\title{Standard Big-Bang Nucleosynthesis after Planck}

\author{Alain Coc}
\affil{Centre de Sciences Nucl\'eaires et de Sciences de la
Mati\`ere  (CSNSM), CNRS/IN2P3, Universit\'e~Paris~Sud~11, UMR~8609,
B\^atiment 104, F--91405 Orsay Campus (France).}

\author{Jean-Philippe Uzan and Elisabeth Vangioni}
\affil{
              (1)Institut d'Astrophysique de Paris,
              UMR-7095 du CNRS, Universit\'e Pierre et Marie
              Curie,\\
              98 bis bd Arago, 75014 Paris (France),\\
              (2) Sorbonne Universit\'es, Institut Lagrange de Paris, 98 bis bd Arago, 75014 Paris (France).}

\begin{abstract}

Primordial or Big Bang nucleosynthesis (BBN)  is one of the three historical strong evidences for the Big-Bang model together with the expansion of the Universe and the Cosmic Microwave Background radiation (CMB). The recent results by the Planck mission have slightly changed the estimate of the  baryonic density $\Omega_{\rm b}$, compared to the  previous WMAP value. This article updates the BBN predictions for the light elements using the new value of $\Omega_{\rm b}$ determined by Planck, as well as an improvement of the nuclear network and new spectroscopic observations. While there is no major modification, the error bars of  the primordial D/H abundance (2.67$\pm$0.09 $ \times10^{-5}$) are narrower and there is a slight lowering of the primordial Li/H abundance (4.89$^{+0.41}_{-0.39}$ $ \times10^{-10}$).   However, this last value is still  $\approx$3 times larger  than its observed spectroscopic abundance in halo stars of the Galaxy. Primordial Helium  abundance is now determined to be $Y_p$ = 0.2463$\pm$0.0003.
\end{abstract}


\today

\section{Introduction}

There are three historical observational evidences for the Big-Bang Model: the cosmic expansion, the Cosmic Microwave Background (CMB) radiation and Primordial or Big-Bang Nucleosynthesis (BBN). Today they are complemented by a large number of evidences in particular from the properties of the large scale structures (see \citet{pubook} for a textbook description). BBN predicts the primordial abundances of the ``light cosmological elements'': \hli\ that are produced during the first 20 mn after the Big-Bang when the Universe was dense and hot enough for nuclear reactions to take place. Comparing the calculated and observed abundances, there is an overall good agreement except for the \sep. The essential cosmological parameter of the model is the baryon to photon ratio, $\eta\equiv n_{\rm b}/n_{\gamma}$ where the photon number density is determined from the CMB temperature and $n_{\rm b}$ is related the baryonic density. $\Omega_{\rm b}$ is now well measured  from the angular power spectrum of the CMB temperature anisotropies. A precise value for this free parameter  was provided by the Wilkinson Microwave Anisotropy Probe (WMAP) satellite, $\Omega_{\rm b}h^2=0.02249\pm0.00056$,~\citep{Kom11} while the recent Planck mission updated it to $\Omega_{\rm b}h^2=0.02207\pm0.00033$~\citep{Planck13}. 

The goal of this letter is to update our previous work~\citep{CV10} to incorporate ($i$) the Planck results, ($ii$) nuclear network improvements and ($iii$) new spectroscopic observations.  To trace the changes since~\citet{CV10}, we follow, step by step,  the effects of each update to identify the key parameters. We first consider the update of the observational data (Section 2) while in Sections~3 and~4, we present the new results with the last nuclear and CMB data and with an extended network respectively.

\section{Primitive observational  abundances update}

Deuterium is a very fragile isotope, easily destroyed after BBN. Its most primitive abundance is determined from the observation of clouds at high redshift, on the line of sight of distant quasars. Very few such  observations  are available. For $\eta=\eta_{\rm WMAP}$, we previously determined a theoretical BBN deuterium abundance, D/H = $(2.59 \pm 0.15) \times 10^{-5}$~\citep{CV10}.

From the observation of about 10  quasar absorption systems  the weighted mean abundance of deuterium is D/H = $(3.02 \pm 0.23) \times 10^{-5}$~\citep{olive2012}.  However, the individual measurements of D/H show a considerable scatter and it is likely that systematic errors dominate the uncertainties. Most of the measurements available in the literature have been gathered in~\citet{pettini08} to deduce D/H~$ = 2.82^{+0.20}_{-0.19}\times 10^{-5}$. Recently, the observation of a Damped Lyman-$\alpha$ (DLA) at $z_{\rm abs}=3.049$ has permitted~\citep{pettini2012}  a new determination of  ${\rm D/H} = 2.535 \pm 0.05 \times 10^{-5}$, leading to a mean determination lower than the previous one, $(2.60 \pm 0.12) \times 10^{-5}$. But, since the H{\sc i} Ly-$\alpha$ absorption associated to this system is redshifted exactly on top of the Ly-$\alpha$-N{\sc v} blend emission from the quasar, the errors on this measurement are probably underestimated. A new analysis is needed to clarify this question and we do not take into account this value presently to determine our weighted mean D abundance. Different star formation histories in the galaxies associated with the DLAs could explain this dispersion. For a recent analysis of the deuterium observations, we refer to \citet{olive2012} and in this present study, we thus adopt their D/H mean abundance value,
\begin{equation}
 {\rm D/H} = 3.02 \pm 0.23 \times 10^{-5}.
\end{equation}

After BBN, \qua\ is still produced by stars, essentially during the main sequence phase. Its primitive abundance is deduced from observations
in H{\sc ii} (ionized hydrogen) regions of compact blue galaxies. In a hierarchical structure formation model, these dwarf galaxies are more primitive than the galaxies. The primordial \qua\ mass fraction, $Y_p$, is obtained from the extrapolation to zero metallicity but is affected by systematic uncertainties \citep{aver10, isot10} such as plasma temperature or stellar absorption. These determinations based on almost the same set of observations lead to
\begin{equation}
Y_p = 0.2561 \pm 0.0108.
\end{equation}

Recently, \citet{aver12} have determined the primordial helium abundance using a Markov Chain Monte Carlo (MCMC) techniques. In this study, a regression to zero metallicity yields 
\begin{equation}
Y_p = 0.2534 \pm 0.0083
\end{equation}

which corresponds to a narrower error bar than previously. We take this last value for comparison with our calculation.

Contrary to \qua\ ,  \tro\ is both produced and destroyed in stars all along its galactic evolution, so that the evolution of its
abundance as a function of time is subject to large uncertainties.
 Moreover, \tro\ has been observed in our Galaxy \citep{Ban02}, and gives only a 'local' constraint
\begin{equation}
\hbox{\tro\ /H}  = 1.1 \pm 0.2 \times 10^{-5}.
\end{equation}
Consequently, the baryometric status of \tro\ is not firmly established \citep{vang03}.

Primitive lithium abundance is deduced from observations of low metallicity stars in the halo of our Galaxy where the lithium abundance is almost independent of metallicity, displaying the so-called Spite plateau \citep{Spite82}. This interpretation assumes that lithium has not been depleted at the surface of these stars, so that the presently observed abundance can be assumed to be equal to the initial one. The small scatter of values around the Spite plateau is indeed an indication that depletion may not have been very effective. However, there is a discrepancy between the value i) deduced from these  observed spectroscopic abundances and ii) the one calculated by BBN  from  \ob CMB observations.  Many studies have been devoted to the resolution of this so-called {\it Lithium problem} and many possible ``solutions'', none fully satisfactory, have been proposed. For a detailed analysis see \citet{fields11} and the proceedings of the meeting ``Lithium in the cosmos''~\citep{LiinC}. 

Astronomical observations of these metal poor halo stars \citep{Ryanetal2000} have thus led to a relative primordial abundance of
${\rm Li/H}= (1.23^{+0.34}_{-0.16}) \times 10^{-10}$ while more recent analysis~\citep{sbordone10} gives
\begin{equation}
  {\rm Li/H}=  (1.58 \pm 0.31) \times 10^{-10}
\end{equation}  
which we use in our analysis. For reviews on the  Li observations, we refer to~\citet{Spite10} and \citet{frebel11}.

\section{New results with nuclear  and CMB data updated}

Since our previous Monte-Carlo BBN calculations \citep{CV10}, no change has been made concerning 11 of the 12 main BBN reactions rates.  We thus use those from the the evaluation performed by  \citet{Des04} except for $^1$H(n,$\gamma$)D \citep{And06} and $^3$He($\alpha,\gamma)^7$Be \citep{Cyb08a}. 

The only modification of one of the main rates concerns the weak reactions involved in n$\leftrightarrow$p equilibrium  whose rates \citep{Dic82} is determined from the standard theory of the weak interaction but needs to be normalized to the experimental neutron  lifetime.  The latter has recently been revised from  885.7$\pm$0.8~s~\citep{PDG08}, used in \citet{CV10}, to 880.1$\pm$1.1~s \citep{PDG12}. This significant change is due to the re-consideration of the previously discarded~\citep{Ser05} experimental value, now comforted by new analyses (see \citet{Wie11,PDG12} for more
details). Comparison between columns 3 and 4 in Table~\ref{t:yields} shows the effect of this change, which remains very small since it lowers $Y_p$ by 0.44\% and \sep/H by 0.39\%, letting the other abundances unchanged.

\begin{table*}[htbp!] 
\caption{\label{t:yields} Primordial abundances. (Bold face displayed values highlight parameter changes.)}
\begin{center}
{\footnotesize\begin{tabular}{ccccccc}
\hline
  &  (a)  &  (b)   & This work  & This work & (b) & This work   \\
 Nb. reactions & 13  & 15 & 15 & 15 & {\bf 424} & 424 \\
\hline
\obh & $0.0223^{+0.00075}_{-0.00073}$ (c) & {\bf 0.02249} (e) & 0.02249 (e) & {\bf 0.02207 (g)}  & 0.02249 (e) & 0.02207{\bf $\pm$0.00033} (g) \\
$\tau_n$ & 885.7$\pm$0.8 (d) & 885.7 (d) & {\bf 880.1 (f)} &  880.1 &  885.7 (d) &  880.1{\bf $\pm$1.1 (f)} \\
\hline
$Y_p$     &  0.2476$\pm$0.0004       & 0.2475    & 0.2464 & 0.24617 & 0.2476  & 0.2463$\pm$0.0003  \\
\deu/H   ($ \times10^{-5})$& $2.68\pm0.15$    & 2.64   &   2.64 & 2.71 &  2.59  & 2.67$\pm$0.09  \\
\tro/H    ($ \times10^{-5}$) & 1.05$\pm$0.04    & 1.05      &   1.05 & 1.06 & 1.04  & 1.05$\pm$0.03       \\
\sep/H ($\times10^{-10}$)  &  5.14$\pm$0.50 &  5.20  & 5.18 & 4.98 & 5.24  & 4.89$^{+0.41}_{-0.39}$  \\
\hline
\end{tabular}}\\
(a) \citet{CV10}; (b) \citet{CGXSV12},
(c) \citet{Spe07} ; (d) \citet{PDG08};\\ (e) \citet{Kom11} ; (f) \citet{PDG12}; (g) \citet{Planck13}
\end{center}
\end{table*}

Concerning the update of the CMB, a comparison between columns  in Table~\ref{t:yields} shows the effect of a change in $\Omega_{\rm b}h^2$\ form
\citet{Spe07} to  \citet{Kom11}  (columns 2, 3) and from \citet{Kom11} to  \citet{Planck13}  (columns 4, 5). It mostly affect \sep/H by about 4\% and D/H by about 2.7\% while the other changes are less than a percent.

A BBN evaluation has been done by \citet{Planck13}, using \obh = 0.02207{$\pm$0.00027};  their prediction regarding the $Y_p$  and \deu/H abundances are similar than ours (0.24725$\pm$0.00032 and   2.656$\pm$0.067 $ \times10^{-5}$ respectively) but they do not provide any \sep/H value.

\section{Extended BBN network and correlated results}

Recently, \citet{CGXSV12} have extended the BBN network by including more than 400 reaction or decay rates in order to
calculate the primordial CNO production during BBN. They performed a sensitivity study by changing each rate within three
orders of magnitudes around their nominal rates or within their known uncertainties when available. 
None of these reactions have displayed a significant influence on the light isotope yields and in particular on 
$^7$Be+\sep\  (see also Hammache et al 2013, submitted to PRC). However the extension of the network with many new neutron producing or absorbing reactions
slightly (-4\%) modify the late neutron abundance, resulting in a moderate increase of the \sep\ yield, as seen
comparing columns 3 and  6 in Table~\ref{t:yields}. 

Concerning this new Monte-Carlo calculation, that involves reactions up to CNO
 \footnote{The results concerning A$>$7 nuclei are beyond the scope of this letter and will be published elsewhere.},
we follow the prescription by \citet{Sal13}. Namely the reaction rates $x$ are assumed to follow a lognormal distribution
with $\mu$ and $\sigma$, tabulated as a function of $T$ and are deduced from the evaluation of rate uncertainties 
by \citet{CGXSV12}. The $p$'s in Eq.~(22)  of \citet{Sal13},  are sampled according
to a normal distribution:
\begin{equation}
x=\exp\left(\mu+p\sigma\right)\equiv x_{\rm med}\left(f.u.\right)^p
\end{equation}
where $x_{\rm med}\equiv\exp\left(\mu\right)$ is the median rate and $f.u.\equiv\exp\left(\sigma\right)$ the factor uncertainty. As discussed in \citet{Eval1}, for small $\sigma$ the lognormal distribution used here is close to a normal distribution as used in \citet{CV10}. (For $\eta$, we use a normal distribution.)
The values displayed in  the last column of Table~\ref{t:yields} correspond to the  0.16, 0.5 and  0.84 quantile of the \hli\ distributions.     

Hence, comparison between columns 2 and 7 in Table~\ref{t:yields} shows the evolution of the yields from \citet{CV10} with the first WMAP results \citep{Spe07} to  the recent Planck data \citep{Planck13}. The reduced uncertainty on D/H is a direct consequence
of the reduced uncertainty on \obh\ while \sep\ uncertainty is still dominated by nuclear uncertainty on the
$^3$He($\alpha,\gamma)^7$Be rate.

\begin{figure}
\begin{center}
\vskip -3.3cm
 \includegraphics[width=.42\textwidth]{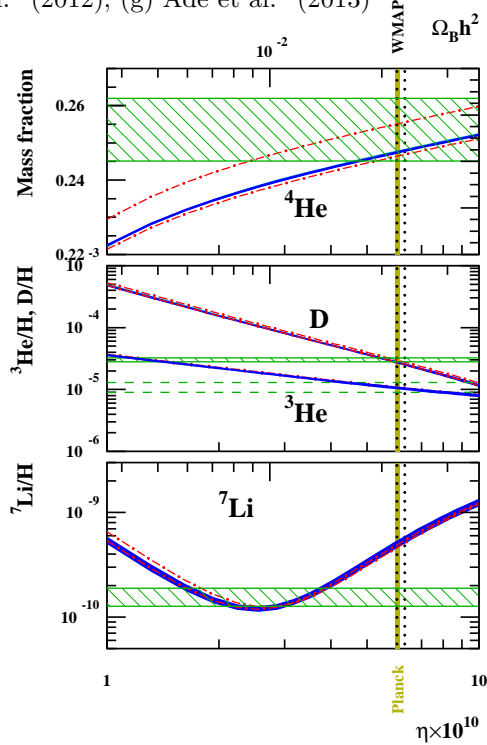}
\begin{minipage}[b]{14pc}
\caption{\label{f:heli} (Color online)
Abundances of \qua\, \deu, \tro\ and \sep\  (blue) as a function of the baryon over photon ratio 
(bottom) or baryonic density (top). The vertical areas corresponds to the WMAP (dot, black) and Planck (solid, yellow) baryonic densities
while the horizontal areas (green) represent the adopted observational abundances; see text.
The  (red) dot--dashed lines correspond to the extreme values of the {\em effective} neutrino families
coming from CMB Planck study, $N_{\rm eff}=(3.02, 3.70)$; see text.}
\end{minipage}
\end{center}
\end{figure}

Figure~\ref{f:heli} displays the abundances as a function of $\eta$ and Table~\ref{t:obs} those at Plank baryonic density, both for $N_{\rm eff}= 3$. 
We do not use the $N_{\rm eff} = 3.046$ value from \citet{mangano05}
to account for non--instantaneous neutrino decoupling in the presence of oscillations.  While this approximation works for \qua, {\em the change for the other
nuclides is exactly in the opposite direction of the true one}. Hence, to implement these very small effects ($\approx2\times10^{-4}$ for $Y_p$), we suggest to
the interested reader,  to correct  $N_{\rm eff}= 3$ results (i.e. Table~\ref{t:obs}) with the exactly calculated abundance changes (e.g. $\Delta Y_p$)
given in the Tables of  \citet{mangano05}, rather than considering $N_{\rm eff} = 3.046$ results. 
In Figure~\ref{f:heli} we also display for visual inspection the results obtained for the limits on effective neutrino family numbers  $N_{\rm eff}= 3.36 \pm0.34$ provided 
by the CMB only confidence interval \citep{Planck13}.  [Note that in \citet{Coc13} we obtained $2.89\leq N_{\rm eff} \leq 4.22$ at WMAP baryonic density,
with $N_{\rm eff}$ defined by eq. 3.12 of the same reference, as in this paper.]
Finally in Table~\ref{t:obs}, a comparison between this work and the last observational data ; an overall consistency between standard BBN calculation and the observational constraints is done except for lithium, as said above: the discrepancy remains of the order of 3.  

\section{Conclusion}

This work has updated the BBN predictions in order to take into account the most recent developments concerning both the cosmological framework (i.e. the cosmological parameters determined from the recent CMB Planck experiment) and the microphysics. It demonstrates that these predictions are robust for the lightest elements. It shows also that the modification of these parameters in the range allowed cannot alleviate the lithium problem.

\begin{table}[htbp!] 
\caption{\label{t:obs} Comparison with observations}
\begin{center}
{\footnotesize\begin{tabular}{ccc}
\hline
   & This work & Observations   \\
\hline
$Y_p$     & 0.2463$\pm$0.0003 & $ 0.2534 \pm 0.0083$ \\
\deu/H   ($ \times10^{-5})$&  2.67$\pm$0.09 & $3.02 \pm 0.23$\\
\tro/H    ($ \times10^{-5}$) &  1.05$\pm$0.03   &  $1.1 \pm 0.2 $  \\
\sep/H ($\times10^{-10}$)  &   4.89$^{+0.41}_{-0.39}$ & $   1.58 \pm 0.31$ \\
\hline
\end{tabular}}\\
\end{center}
\end{table}

\section*{Acknowledgements}
This work made in the ILP LABEX (under reference ANR-10-LABX-63) was supported by French state funds managed by the ANR 
within the Investissements d'Avenir programme under reference ANR-11-IDEX-0004-02 and by the ANR VACOUL, ANR-10-BLAN-0510.

\end{document}